\documentclass[11pt]{article}%
\usepackage{amsfonts}
\usepackage{amsmath}%
\usepackage{amssymb}%
\usepackage{graphicx}
\newtheorem{theorem}{Theorem}

\begin{document}

\title{Systematic construction of separable systems with quadratic in momenta first integrals}
\author{Maciej B\l aszak\\Institute of Physics, A. Mickiewicz University\\Umultowska 85, 61-614 Pozna\'{n}, Poland}
\maketitle

\begin{abstract}
Liouville integrable separable systems with quadratic in momenta first
integrals are considered. Particular attention is paid to the systems
generated by the so-called special conformal Killing tensors, i.e. Benenti
systems. Then, infinitely many new classes of separable systems are
constructed by appropriate deformations of Benenti class systems.

\end{abstract}

\section{Introduction}

The separation of variables for solving by quadratures the Hamilton-Jacobi
(HJ) equations of related Liouville integrable dynamic systems with quadratic
in momenta first integrals has a long history as a part of analytical
mechanics. There are some mile stones of that theory. First, in 1891
St\"{a}ckel initiated a program of classification of separable systems
presenting conditions for separability of the HJ equations in orthogonal
coordinates \cite{Blaszak:s1}-\cite{Blaszak:s3}. Then, in 1904 Levi-Civita
found a test for the separability of a Hamiltonian dynamics in a given system
of canonical coordinates \cite{Blaszak:l}. The next was Eisenhart
\cite{Blaszak:Ei}-\cite{Blaszak:Ei1}, who in 1934 inserted a separability
theory in the context of Riemannian geometry, making it coordinate free and
introducing the crucial objects of the theory, i.e. Killing tensors. This
approach was then developed by Woodhouse \cite{Blaszak:woo}, Klanins
\cite{Blaszak:kl}-\cite{Blaszak:KL1} and others. Finally, in 1992, Benenti
\cite{Blaszak:be} -\cite{Blaszak:be1} constructed a particular but very
important subclass of separable systems, based on the so called special
conformal Killing tensors.

The first constructive theory of separated coordinates for dynamical systems
was made by Sklyanin \cite{Blaszak:sk1}. He adopted the method of soliton
theory, i.e. the Lax representation, to systematic derivation of separated
coordinates. In that approach involutive functions appear as coefficients of
characteristic equation (spectral curve) of Lax matrix. The method was
successfully aplied to separation of variables for many integrable systems
\cite{Blaszak:sk1}-\cite{Blaszak:ma1}.

Recently, a modern geometric theory of separability on bi-Poisson manifolds
was constructed \cite{Blaszak:1}-\cite{Blaszak:m3}, related to the so-called
Gel'fand-Zakharevich (GZ) bi-Hamiltonian systems \cite{Blaszak:GZ}%
,\cite{Blaszak:GZ1}. \ Obviously, it contains as a special case Liouville
integrable systems with all constants of motion being quadratic in momenta
functions. Indeed, Ibort et.al. \cite{Blaszak:ib} proved that the Benenti
class of systems can be lifted to the GZ bi-Hamiltonian form.

In the following paper we construct in a systematic way all Liouville
integrable systems on Riemannian spaces, which are of the GZ type, including
as a special case the Benenti class of systems. What is important, infinitely
many classes of separable systems are constructed from appropriate
deformations of the Benenti class of systems. In that sense we demonstrate the
crucial role of this particular class of systems in the separability theory of
dynamic systems with quadratic in momenta first integrals.

\section{Separable dynamics on Riemannian spaces}

Let $(Q,g)$ be a Riemann (pseudo-Riemann) manifold with covariant metric
tensor $g$ and local coordinates $q^{1},...,q^{n}.$ Moreover, let $G:=g^{-1}$
be a contravariant metric tensor satisfying $\sum_{j=1}^{n}g_{ij}G^{jk}%
=\delta_{i}^{k}.$ The equations
\begin{equation}
q_{tt}^{i}+\Gamma_{jk}^{i}q_{t}^{j}q_{t}^{k}=G^{ik}\partial_{k}%
V(q),\ \ \ \ \ \ \ \ i=1,...,n,\ \ \ \ \ \ \ q_{t}\equiv\frac{dq}%
{dt}\label{Blaszak:1.2}%
\end{equation}
describe the motion of a particle in the Riemannian space with the metric $g$,
where $\Gamma_{jk}^{i}$ are the Levi-Civita connection components$.$
Eqs.(\ref{Blaszak:1.2}) can be obtained by varying the Lagrangian
\begin{equation}
\mathcal{L}(q,q_{t})=\frac{1}{2}\sum_{i,j}g_{ij}q_{t}^{i}q_{t}^{j}%
-V(q)\label{Blaszak:1.3}%
\end{equation}
and are called Euler-Lagrange equations. Obviously, for $G=I$ equations
(\ref{Blaszak:1.2}) reduce to Newton equations of motion.

One can pass in a standard way to the Hamiltonian description of dynamics,
where the Hamiltonian function takes the form
\[
H(q,p)=\sum_{i=1}^{n}q_{t}^{i}\frac{\partial\mathcal{L}}{\partial q_{t}^{i}%
}-\mathcal{L}=\frac{1}{2}\sum_{i,j=1}^{n}G^{ij}p_{i}p_{j}+V(q),\ \ \ \ \ p_{i}%
:=\frac{\partial\mathcal{L}}{\partial q_{t}^{i}}=\sum_{j}g_{ij}q_{t}^{j}%
\]
and equations of motion are
\[
\left(
\begin{array}
[c]{c}%
q\\
p
\end{array}
\right)  _{t}=\theta_{0}dH=\left(
\begin{array}
[c]{cc}%
0 & I\\
-I & 0
\end{array}
\right)  \left(
\begin{array}
[c]{c}%
\frac{\partial H}{\partial q}\\
\frac{\partial H}{\partial\partial p}%
\end{array}
\right)  =X_{H}\ \Longleftrightarrow q_{t}^{i}=\frac{\partial H}{\partial
p_{i}},\ \ \ \ p_{it}=-\frac{\partial H}{\partial q^{i}}.
\]
$X_{H}$ denotes the Hamiltonian vector field with respect to a canonical
Poisson tensor $\theta_{0}$ and the whole dynamics takes place on the phase
space $\mathcal{N}=T^{\ast}G$ in local coordinates $(q^{1},...,q^{n}%
,p_{1},...,p_{n}).$

Of special importance is the geodesic motion $V(q)\equiv0,$ with
Euler-Lagrange equations and Hamiltonian representation in the form
\[
q_{tt}^{i}+\Gamma_{jk}^{i}q_{t}^{j}q_{t}^{k}=0,\ \ \ \ i=1,...,n\ \ \
\]%
\[
\Updownarrow
\]%
\[
\ \left(
\begin{array}
[c]{c}%
q\\
p
\end{array}
\right)  _{t}=\theta_{0}dE=X_{E},\ \ E=\frac{1}{2}\sum_{i,j=1}^{n}G^{ij}%
p_{i}p_{j}.
\]

Functionally independent Hamiltonian functions $H_{i},i=1,...,n$ are said to
be separable in the canonical coordinates $(\lambda,\mu)$ if there are $n$
relations, called the separation conditions (Sklyanin \cite{Blaszak:sk1}), of
the form
\begin{equation}
\varphi_{i}(\lambda^{i},\mu_{i};H_{1},...,H_{n})=0,\;\;\;i=1,...,n,\ \ \det
\left[  \frac{\partial\varphi_{i}}{\partial H_{j}}\right]  \neq
0,\label{Blaszak:0.17}%
\end{equation}
which guarantee the solvability of the appropriate Hamilton-Jacobi equations
and involutivity of $H_{i}$. A special case, when all separation relations
(\ref{Blaszak:0.17}) are affine in $H_{i},$ is given by the set of equations
\begin{equation}
\sum_{k=1}^{n}\phi_{i}^{k}(\lambda_{i},\mu_{i})H_{k}=\psi_{i}(\lambda_{i}%
,\mu_{i}),\;\;\;i=1,...,n,\label{Blaszak:0.18}%
\end{equation}
where $\phi$ and $\psi$ are arbitrary smooth functions of their arguments, is
called the \emph{St\"{a}ckel separation conditions} and the related dynamic
systems are called St\"{a}ckel separable.

We are going to present a subclass of one-particle dynamics, containing
Liouville integrable and separable systems with $n$ quadratic in momenta
constants of motion. St\"{a}ckel \cite{Blaszak:s1}-\cite{Blaszak:s3} was a
first who gave the characterization of equations of motion integrable by
separation of variables. He proved that if in a system of orthogonal
coordinates $(\lambda,\mu)$ there exists a non-singular matrix $\varphi
=(\varphi_{k}^{l}(\lambda_{k}))$, called a \emph{St\"{a}ckel matrix} such that
the Hamiltonians $H_{r}$ are of the form
\begin{equation}
H_{r}=\frac{1}{2}\sum_{i=1}^{n}(\varphi^{-1})_{r}^{i}(\mu_{i}^{2}+\sigma
_{i}(\lambda_{i})),\label{Blaszak:a}%
\end{equation}
then $H_{r}$ are functionally independent, pairwise commute with respect to
the canonical Poisson bracket and the Hamilton-Jacobi equation associated to
$H_{1}$ is separable. Indeed, for quadratic in momenta constants of motion,
the St\"{a}ckel separation conditions (\ref{Blaszak:0.18}) take the general
form
\begin{equation}
\sum_{k=1}^{n}\phi_{i}^{k}(\lambda^{i})H_{k}=\frac{1}{2}f_{i}(\lambda^{i}%
)\mu_{i}^{2}+\gamma_{i}(\lambda^{i})\;\;\;i=1,...,n,\label{Blaszak:13}%
\end{equation}
where $f_{i},\gamma_{i},\phi_{i}^{k}$ are arbitrary smooth functions of its
argument and the normalization $\phi_{i}^{n}=1,\;i=1,...,n$ is assumed. To get
the explicit form of $H_{k}=H_{k}(\lambda,\mu)$ one has to solve the system of
linear equations (\ref{Blaszak:13}). The results are the following%
\[
\varphi=\left(
\begin{array}
[c]{ccccc}%
\frac{\phi_{1}^{1}(\lambda^{1})}{f_{1}(\lambda^{1})} & \frac{\phi_{1}%
^{2}(\lambda^{1})}{f_{1}(\lambda^{1})} & \cdots & \frac{\phi_{1}^{n-1}%
(\lambda^{1})}{f_{1}(\lambda^{1})} & \frac{1}{f_{1}(\lambda^{1})}\\
\vdots & \vdots & \cdots & \vdots & \vdots\\
\frac{\phi_{n}^{1}(\lambda^{n})}{f_{n}(\lambda^{n})} & \frac{\phi_{n}%
^{2}(\lambda^{n})}{f_{n}(\lambda^{n})} & \cdots & \frac{\phi_{n}^{n-1}%
(\lambda^{n})}{f_{n}(\lambda^{n})} & \frac{1}{f_{n}(\lambda^{n})}%
\end{array}
\right)  ,\ \ \ \ \sigma_{i}(\lambda_{i})=\gamma_{i}(\lambda^{i}%
)/f_{i}(\lambda^{i}).
\]

Eisenhart considered St\"{a}ckel separable systems in the frame of
one-particle dynamics on Riemannian (pseudo-Riemannian) space $(Q,g)$. He gave
a coordinate-free representation for St\"{a}ckel geodesic motion introducing a
special family of \emph{Killing tensors }\cite{Blaszak:Ei}-\cite{Blaszak:Ei1}.
As known, a $(1,1)$-type tensor $K=(K_{j}^{i})$ (or a $(2,0)$-type tensor
$KG=A=(A^{ij}))$ is called a Killing tensor with respect to $g$ if
\[
\left\{  \,\sum A^{ij}p_{i}p_{j}\,,\sum G^{ij}p_{i}p_{j}\right\}  _{\theta
_{0}}=0,
\]
where $\{.,.\}_{\theta_{0}}$ means a canonical Poisson bracket. He proved
\cite{Blaszak:Ei}-\cite{Blaszak:Ei1} that the geodesic Hamiltonians can be
transformed into the St\"{a}ckel form (\ref{Blaszak:a}) if the contravariant
metric tensor $G=g^{-1}$ has $(n-1)$ commuting independent contravariant
Killing tensors $A_{r}$ of a second order such that
\[
E_{r}=\frac{1}{2}\sum_{i,j}A_{r}^{ij}p_{i}p_{j},
\]
admitting a common system of closed eigenforms $\alpha_{i}$
\[
(A_{r}^{\ast}-v_{r}^{i}G)\alpha_{i}=0,\;\;d\alpha_{i}=0,\;\;i=1,...,n,
\]
where $v_{r}^{i}$ are eigenvalues of $(1,1)$ Killing tensor $K_{r}=A_{r}g$
$(K_{r}^{\ast}=gA_{r}^{\ast}).$

For $n$ degrees of freedom, let us consider $n$ St\"{a}ckel Hamiltonian
functions in separated coordinates in the following form
\begin{equation}
H_{r}=\frac{1}{2}\sum_{i=1}^{n}v_{r}^{i}G^{ii}\mu_{i}^{2}+V_{r}(\lambda
)=\frac{1}{2}\mu^{T}K_{r}G\mu+V_{r}(\lambda),\;\;\;r=1,...,n,\label{Blaszak:6}%
\end{equation}
where $\mu=(\mu_{1},...,\mu_{n})^{T}$and $V_{r}(\lambda)$ are appropriate
potentials separable in $(\lambda,\mu)$ coordinates. For the integrable system
(\ref{Blaszak:6}) calculated from (\ref{Blaszak:13})
\[
G^{ii}=(-1)^{i+1}\frac{f_{i}(\lambda^{i})\det W^{i1}}{\det W},\;\;\;v_{r}%
^{i}=(-1)^{r+1}\frac{\det W^{ir}}{\det W^{i1}},
\]%
\[
V_{r}=\sum_{i=1}^{n}(-1)^{i+r}\gamma_{i}(\lambda^{i})\frac{\det W^{ir}}{\det
W},
\]
where
\[
W=\left(
\begin{array}
[c]{ccccc}%
\phi_{1}^{1}(\lambda^{1}) & \phi_{1}^{2}(\lambda^{1}) & \cdots & \phi
_{1}^{n-1}(\lambda^{1}) & 1\\
\vdots & \vdots & \cdots & \vdots & \vdots\\
\phi_{n}^{1}(\lambda^{n}) & \phi_{n}^{2}(\lambda^{n}) & \cdots & \phi
_{n}^{n-1}(\lambda^{n}) & 1
\end{array}
\right)
\]
and $W^{ik}$ is the $(n-1)\times(n-1)$ matrix obtained from $W$ after we
cancel its $i$th row and $k$th column.

In our further considerations we restrict to the so-called GZ case, when
$\phi_{i}^{k}(\lambda_{i})$ are monomials, $f_{i}=f$ and $\gamma_{i}=\gamma$.
Then, the St\"{a}ckel separation conditions are $n$ copies of a so-called
separation curve
\begin{equation}
H_{1}\xi^{m_{1}}+...+H_{n}\xi^{m_{n}}=\frac{1}{2}f(\xi)\mu^{2}+\gamma
(\xi),\ \ \ m_{n}=0<m_{n-1}<...<m_{1}\in\mathbb{N},\label{Blaszak:stackel}%
\end{equation}
with $\ (\xi,\mu)=(\lambda^{i},\mu_{i}),\ i=1,...,n.$

In $(\lambda,\mu)$ coordinates $n$ related Hamilton-Jacobi (HJ) equations
\[
H_{r}(\lambda,\frac{\partial W}{\partial\lambda})=a_{r},\;\;\;\;\;r=1,...,n,
\]
for a generation function $W(\lambda,a)=\sum_{i=1}^{n}W_{i}(\lambda_{i},a)$,
decouples into $n$ ordinary differential equations
\[
\frac{1}{2}f(\lambda^{i})\left(  \frac{dW_{i}}{d\lambda^{i}}\right)
^{2}+\gamma(\lambda^{i})=a_{1}(\lambda^{i})^{m_{1}}+a_{2}(\lambda^{i})^{m_{2}%
}+...+a_{n}\equiv a(\lambda^{i})
\]
and hence, the implicit solution of dynamical system with Hamiltonian function
$H_{r}$ is given by
\[
\sum_{k=1}^{n}\int^{\lambda^{k}}\frac{\xi^{m_{i}}}{\sqrt{\psi(\xi)}}d\xi
=t_{r}\delta_{ri}+const_{i},\ \ \ \ \ \ i=1,...,n,
\]
where $2f(\xi)[a(\xi)-\gamma(\xi)]\equiv\psi(\xi),$ called the \emph{inverse
Jacobi problem}.

In this context, a question about classification and construction in natural
coordinates of all separable systems on Riemannian spaces, with $n$ quadratic
in momenta constants of motion, arises. The classification can be made with
respect to the admissible forms of St\"{a}ckel separability conditions. The
right hand side of the conditions (\ref{Blaszak:stackel}) is always the same
for the class of systems considered
\begin{equation}
r.h.s.=\frac{1}{2}f(\lambda^{i})\mu_{i}^{2}+\gamma(\lambda^{i})=\psi
(\lambda^{i},\mu_{i}),\label{Blaszak:3.1}%
\end{equation}
so different classes of separable systems are described by different forms of
the l.h.s. of St\"{a}ckel conditions, while systems from the same class are
described by different $f$ and $\gamma$ in (\ref{Blaszak:3.1}).

\section{Separable systems in natural coordinates}

Among all St\"{a}ckel systems a particularly important subclass consists of
these considered by Benenti \cite{Blaszak:be}, \cite{Blaszak:ben},
\cite{Blaszak:be1} and constructed with the help of the so-called
\emph{special conformal Killing tensor. }Let $L=(L_{j}^{i})$ be a second order
mixed type tensor on $Q$ and let $\overline{L}:M\rightarrow\mathbb{R}$ be a
function on $M$ defined as $\overline{L}=\frac{1}{2}\sum_{i,j=1}^{n}%
(LG)^{ij}p_{i}p_{j}$. If
\[
\{\overline{L},E\}_{\theta_{0}}=\kappa E,\ \ \ \ \mathrm{\ where}%
\ \ \ \ \ \ \kappa=\{\varepsilon,E\}_{\theta_{0}},\ \ \ \ \varepsilon=Tr(L),
\]
then $L$ is called a special conformal Killing tensor with the associated
potential $\varepsilon=Tr(L)$. An important property of $L$ is the vanishing
of its Nijenhuis torsion.

For the Riemannian space $(Q,g,L),$ geodesic flow has $n$ constants of motion
of the form
\begin{equation}
E_{r}=\frac{1}{2}\sum_{i,j=1}^{n}A_{r}^{ij}p_{i}p_{j}=\frac{1}{2}\sum
_{i,j=1}^{n}(K_{r}G)^{ij}p_{i}p_{j}%
,\ \ \ \ \ \ \ \ \ \ r=1,...,n,\label{Blaszak:3.4}%
\end{equation}
where $A_{r}$ and $K_{r}$ are Killing tensors of type $(2,0)$ and $(1,1)$,
respectively. Moreover, as was shown by Benenti \cite{Blaszak:be}%
,\cite{Blaszak:ben}, all the Killing tensors $K_{r}$ with a common set of
eigenvectors, are constructed from $L$ by
\[
K_{r+1}=\sum_{k=0}^{r}\rho_{k}L^{r-k},
\]
where $\rho_{r}$ are coefficients of the characteristic polynomial of $L$%
\begin{equation}
\det(\xi I-L)=\xi^{n}+\rho_{1}\xi^{n-1}+...+\rho_{n},\ \ \ \rho_{0}%
=1,\label{Blaszak:2.8b}%
\end{equation}
or equivalently by the following 'cofactor' formula \cite{Blaszak:5}%
,\cite{Blaszak:pre}
\[
cof(\xi I-L)=\sum_{i=0}^{n-1}K_{n-i}\xi^{i},
\]
where $cof(A)$ stands for the matrix of cofactors, so that $cof(A)A=(\det
A)I.$ So, for a given metric tensor $g$, the existence of a special conformal
Killing tensor $L$ is a sufficient condition for the geodesic flow on $Q$ to
be a Liouville integrable Hamiltonian system with all constants of motion
quadratic in momenta. Moreover, the basic separable potentials $V_{r}^{(m)}$
are given by the following recursion relation \cite{Blaszak:5},
\cite{Blaszak:8}%
\begin{equation}
V_{r}^{(m+1)}=\rho_{r}V_{1}^{(m)}-V_{r+1}^{(m)},\label{Blaszak:pot}%
\end{equation}
where the first nontrivial potentials are $V_{r}^{(0)}=-\rho_{r}.$

It turns out that with the tensor $L$ we can (generically) associate a
coordinate system on $Q$ in which the geodesic flows associated with all the
functions $E_{r}$ separate. Namely, let $(\lambda^{1}(q),...,\lambda^{n}(q))$
be $n$ distinct, functionally independent eigenvalues of $L$, i.e. solutions
of the characteristic equation \ $\det(\xi I-L)=0$. Solving these relations
with respect to $q$ we get the transformation $\lambda\rightarrow
q:\ q_{i}=\alpha_{i}(\lambda)$. The remaining part of the transformation to
the separation coordinates can be reconstructed from the generating function
$W(p,\lambda)=\sum_{i}p_{i}\alpha_{i}(\lambda).$ In the $(\lambda,\mu)$
coordinates the St\"{a}ckel separation conditions (\ref{Blaszak:stackel}) for
Benenti Hamiltonian functions $H_{r}=E_{r}+V_{r}^{(n-1+j)}$ are given by the
separation curve of the form
\begin{equation}
H_{1}\xi^{n-1}+H_{2}\xi^{n-2}+...+H_{n}=\frac{1}{2}f(\xi)\mu^{2}+\xi
^{n+j},\ \ \ \ j=0,1,2,...\ .\label{Blaszak:2.24}%
\end{equation}

It is important to notice, that all Liouville integrable systems of classical
mechanics, with quadratic in momenta first integrals, that was separated in
XIX and XX centuries, belong to the Benenti class. The reason is that only the
Benenti class contains dynamic systems on flat and constant curvature
Riemannian spaces. Another important feature of Benenti systems is that all of
them can be lifted to a one Casimir bi-Hamiltonian form \cite{Blaszak:ib}.

Now we present how to construct all remaining classes of separable systems by
an appropriate deformations of the Benenti class. Let start from the
separability condition (\ref{Blaszak:stackel}) for $n$ Hamiltonian functions
in the following form
\begin{equation}
\widetilde{H}_{1}\xi^{(n+k)-1}+\widetilde{H}_{2}\xi^{(n+k)-2}+...+\widetilde
{H}_{n+k}=\frac{1}{2}f(\xi)\mu^{2}+\xi^{n+j},\ \ \ \ \ \ k\in\mathbb{N,}%
\ \ j\geq k,\label{Blaszak:6.1}%
\end{equation}
\ where $\widetilde{H}_{n_{1}}=\widetilde{H}_{n_{2}}=...=\widetilde{H}_{n_{k}%
}=0$, $1<n_{1}<...<n_{k}<n+k,$ and the separability condition for Benenti
systems with the same $\psi$%
\begin{equation}
H_{1}\xi^{n-1}+H_{2}\xi^{n-2}+...+H_{n}=\frac{1}{2}f(\xi)\mu^{2}+\xi
^{n+j}.\label{Blaszak:6.2}%
\end{equation}

\begin{theorem}
Deformation of the Benenti system with separation curve (\ref{Blaszak:6.2}) to
the new system with separation curve (\ref{Blaszak:6.1}) is given by a
following determinant form
\begin{equation}
\widetilde{H}_{r}=\frac{\left\vert
\begin{array}
[c]{cccc}%
H_{r-k} & \rho_{r-1} & \cdots & \rho_{r-k}\\
H_{n_{1}-k} & \rho_{n_{1}-1} & \cdots & \rho_{n_{1}-k}\\
\cdots & \cdots & \cdots & \cdots\\
H_{n_{k}-k} & \rho_{n_{k}-1} & \cdots & \rho_{n_{k}-k}%
\end{array}
\right\vert }{\left\vert
\begin{array}
[c]{ccc}%
\rho_{n_{1}-1} & \cdots & \rho_{n_{1}-k}\\
\cdots & \cdots & \cdots\\
\rho_{n_{k}-1} & \cdots & \rho_{n_{k}-k}%
\end{array}
\right\vert },\label{Blaszak:6.4}%
\end{equation}
where $\rho_{i},\ i=0,...,n$ are coefficients of the characteristic polynomial
of the conformal Killing tensor $L$ (\ref{Blaszak:2.8b}) related to the
Benenti system.
\end{theorem}

The formula (\ref{Blaszak:6.4}) applies separately to the geodesic and the
potential parts. The geodesic part $\widetilde{E}_{r}$ can be presented in the
following form
\begin{equation}
\widetilde{E}_{r}=\frac{1}{2}p^{T}\widetilde{K}_{r}\widetilde{G}%
p,\ \ \ \ \ r=1,...,n+k,\label{Blaszak:10}%
\end{equation}
where metric tensor $\widetilde{G}$ and Killing tensors $\widetilde{K}_{r}$
are
\[
\ \widetilde{G}=(-1)^{k}\frac{1}{\varphi}D_{0}G,\ \ \ \widetilde{K}_{r}%
=K_{r}-K_{r-1}D_{1}D_{0}^{-1}+...+(-1)^{k}K_{r-k}D_{k}D_{0}^{-1},
\]%
\[
\varphi=\left\vert
\begin{array}
[c]{ccc}%
\rho_{n_{1}-1} & \cdots & \rho_{n_{1}-k}\\
\cdots & \cdots & \cdots\\
\rho_{n_{k}-1} & \cdots & \rho_{n_{k}-k}%
\end{array}
\right\vert ,\ \ \ D_{0}=\left\vert
\begin{array}
[c]{ccc}%
K_{n_{1}-1} & \cdots & K_{n_{1}-k}\\
\cdots & \cdots & \cdots\\
K_{n_{k}-1} & \cdots & K_{n_{k}-k}%
\end{array}
\right\vert ,
\]%
\[
D_{i}=\left\vert
\begin{array}
[c]{cccccc}%
K_{n_{1}} & \cdots & K_{n_{1}-i+1} & K_{n_{1}-i-1} & \cdots & K_{n_{1}-k+1}\\
\cdots & \cdots & \cdots & \cdots & \cdots & \cdots\\
K_{n_{k}} & \cdots & K_{n_{k}-i+1} & K_{n_{k}-i-1} & \cdots & K_{n_{k}-k+1}%
\end{array}
\right\vert ,\ \ i=1,...,k
\]
and $K_{m}$ in determinant calculations are treated as symbols not matrices.
The proof of the theorem as well as other details the reader can find in
\cite{Blaszak:pre}.

Such constructed systems, although obtained through the deformation procedure
on the level of Hamiltonian functions, are far from being trivial
generalizations of Benenti systems. There is no obvious relations between
solutions of a given Benenti system and all its deformations. In each case we
have a different inverse Jacobi problem to solve. Notice, that the common
feature of appropriate deformed systems is the same set of separated
coordinates, determined by the related Benenti system. Moreover all of them
can be lifted to multi-Casimir bi-Hamiltonian form \cite{Blaszak:pre}.

\section{Example}

Consider the case $n=2.$ Let $Q$ be\ a two dimensional flat space parametrized
by canonical coordinates $q=(q^{1},q^{2})$ with the contravariant metric
tensor $G$ and related special conformal Killing tensor $L$ of the form
\[
G=\left(
\begin{array}
[c]{cc}%
1 & 0\\
0 & 1
\end{array}
\right)  ,\ L=\left(
\begin{array}
[c]{cc}%
q^{1} & \frac{1}{2}q^{2}\\
\frac{1}{2}q^{2} & 0
\end{array}
\right)  \Longrightarrow K_{2}=\left(
\begin{array}
[c]{cc}%
0 & \frac{1}{2}q^{2}\\
\frac{1}{2}q^{2} & -q^{1}%
\end{array}
\right)  .
\]
Two geodesic Hamiltonians $E_{1}$ and $E_{2}$, according to (\ref{Blaszak:3.4}%
) are%
\[
E_{1}=\frac{1}{2}p_{1}^{2}+\frac{1}{2}p_{2}^{2},\ \ \ \ \ E_{2}=\frac{1}%
{2}q^{2}p_{1}p_{2}-\frac{1}{2}q^{1}p_{2}^{2}.
\]
Let us choose the potential $V^{(2)}$. As $\rho_{1}=-q^{1}$ and $\rho
_{2}=-\frac{1}{4}(q^{2})^{2}$ hence
\[
V_{1}^{(2)}=(q^{1})^{3}+\frac{1}{2}q^{1}(q^{2})^{2},\ \ \ \ \ V_{2}%
^{(2)}=\frac{1}{16}(q^{2})^{4}+\frac{1}{4}(q^{1})^{2}(q^{2})^{2}.
\]
It is one of the integrable cases of the Henon-Heiles system with Hamiltonian
function $H_{1}=E_{1}+$ $V_{1}^{(3)}$, second constant of motion $H_{2}%
=E_{2}+$ $V_{2}^{(3)}$ and Newton equations
\begin{align*}
(q^{1})_{tt}  & =-3(q^{1})^{2}-\frac{1}{2}(q^{2})^{2}\\
(q^{2})_{tt}  & =-q^{1}q^{2}.
\end{align*}

The transformation to separated coordinates $(\lambda,\mu)$ takes the form
\[
q^{1}=\lambda^{1}+\lambda^{2},\ \ \ q^{2}=2\sqrt{-\lambda^{1}\lambda^{2}},
\]%
\[
p_{1}=\frac{\lambda^{1}\mu_{1}}{\lambda^{1}-\lambda^{2}}+\frac{\lambda^{2}%
\mu_{2}}{\lambda^{2}-\lambda^{1}},\ \ \ p_{2}=\sqrt{-\lambda^{1}\lambda^{2}%
}\left(  \frac{\mu_{1}}{\lambda^{1}-\lambda^{2}}+\frac{\mu_{2}}{\lambda
^{2}-\lambda^{1}}\right)  ,
\]
and the separation curve is
\[
H_{1}\xi+H_{2}=\frac{1}{2}\xi\mu^{2}+\xi^{4}.
\]

Now, let us consider the simplest deformation given by $k=1,\ n_{1}=2.$ Then,
\[
\widetilde{G}=-\frac{1}{\rho_{1}}G=\left(
\begin{array}
[c]{cc}%
\frac{1}{q^{1}} & 0\\
0 & \frac{1}{q^{1}}%
\end{array}
\right)  ,\widetilde{K}_{2}=0,\widetilde{K}_{3}=-K_{2}^{2}=\left(
\begin{array}
[c]{cc}%
-\frac{1}{4}(q^{2})^{2} & \frac{1}{2}q^{1}q^{2}\\
\frac{1}{2}q^{1}q^{2} & -\frac{1}{4}(q^{2})^{2}-(q^{1})^{2}%
\end{array}
\right)  ,
\]%
\[
\widetilde{V}_{1}=-\frac{1}{\rho_{1}}V_{1},\ \ \widetilde{V}_{2}%
=0,\ \ \widetilde{V}_{3}=V_{2}-\frac{\rho_{2}}{\rho_{1}}V_{1},
\]
hence
\begin{align*}
\widetilde{H}_{1}  & =\frac{1}{2}\frac{1}{q^{1}}p_{1}^{2}+\frac{1}{2}\frac
{1}{q^{1}}p_{2}^{2}+(q^{1})^{2}+\frac{1}{2}(q^{2})^{2},\\
\widetilde{H}_{3}  & =-\frac{1}{8}\frac{q_{2}^{2}}{q^{1}}p_{1}^{2}+\frac{1}%
{2}q^{2}p_{1}p_{2}-\frac{1}{8}\frac{q_{2}^{2}}{q^{1}}p_{2}^{2}-\frac{1}%
{16}(q^{2})^{4}.
\end{align*}
$\widetilde{H}_{1}$ and $\widetilde{H}_{2}$ are in involution and are
separated in the same coordinates $(\lambda,\mu)$ as the Henon-Heiles system.
The appropriate separation curve takes the form
\[
\widetilde{H}_{1}\xi^{2}+\widetilde{H}_{3}=\frac{1}{2}\xi\mu^{2}+\xi^{4}.
\]

\end{document}